\documentclass{article}
\usepackage[preprint]{spconf}
\usepackage{hyperref}       % hyperlinks
\usepackage{url}   % simple URL typesetting
\usepackage{soul}
\usepackage{booktabs}       % professional-quality tables
\usepackage{amsfonts}       % blackboard math symbols
\usepackage{nicefrac}       % compact symbols for 1/2, etc.
\usepackage{microtype}      % microtypography
\usepackage{xcolor}         % colors
\usepackage{subcaption}
\usepackage{graphicx}
\usepackage{balance}
\usepackage{float}
\usepackage{amsmath}
\usepackage{enumitem}
\usepackage{multirow}
\title{Revelio: A Real-World Screen-Camera Communication System with Visually Imperceptible Data Embedding}
\name{Abbaas Alif Mohamed Nishar$^{\star}$ \qquad Shrinivas Kudekar$^{\dagger}$ \qquad Bernard Kintzing$^{\ddagger}$ \qquad Ashwin Ashok$^{\star}$}

\address{$^{\star}$ Georgia State University, Atlanta, Georgia, USA \\
         $^{\dagger}$ Revelio Communications Inc., Atlanta, Georgia, USA \\
         $^{\ddagger}$ Bernard Kintzing LLC, Helena, Montana, USA \\
         \small{\texttt{amohamednishar1@student.gsu.edu}, \texttt{skudekar@revelio.ai}, \texttt{bernardkintzing@gmail.com}, \texttt{aashok@gsu.edu}}}

\begin{document}
\maketitle
\noindent{\footnotesize \textcopyright\ 2025 IEEE. Personal use of this material is permitted. Permission from IEEE must be obtained for all other uses, \\
in any current or future media, including reprinting/republishing this material for advertising or promotional purposes, \\
creating new collective works, for resale or redistribution to servers or lists, or reuse of any copyrighted component of this work in other works.}
% \setcopyright{acmcopyright}
% \copyrightyear{2024}
% \acmYear{2024}
% \acmDOI{xx.xxxx/xxxxxxx.xxxxxxx}

% %% These commands are for a PROCEEDINGS abstract or paper.
% \acmConference[EWSN '24]{}{December 10 - December 13, 2024}{Abu Dhabi, UAE}
% %
% %  Uncomment \acmBooktitle if the title of the proceedings is different
% %  from ``Proceedings of ...''!
% %
% %\acmBooktitle{Woodstock '18: ACM Symposium on Neural Gaze Detection,
% %  June 03--05, 2018, Woodstock, NY} 
% \acmPrice{15.00}
% \acmISBN{xxx-x-xxxx--xxxx-12/24/10}

% \author{Abbaas Alif Mohamed Nishar}
% %\authornote{First author}
% \authornotemark[1]
% \affiliation{%
%   \institution{Georgia State University}
%   \city{Atlanta}
%   \state{Georgia}
%   \country{USA}
%   \postcode{30303}}
% \email{amohamednishar1@student.gsu.edu}

% \author{Shrinivas Kudekar}
% \authornotemark[1]
% \affiliation{
% \institution{Revelio Communications Inc.}
% \city{Atlanta}
% \state{Georgia}
% \country{USA}
% \postcode{30075}
% }
% \email{skudekar@revelio.ai}

% \author{Bernard Kintzing}
% \affiliation{
% \institution{Bernard Kintzing LLC}
% \city{Helena}
% \state{Montana}
% \country{USA}
% \postcode{5900601}
% }
% \email{bernardkintzing@gmail.com}
% \author{Ashwin Ashok}
% \affiliation{%
%   \institution{Georgia State University}
%   \city{Atlanta}
%   \state{Georgia}
%   \country{USA}
%   \postcode{30303}}
% \email{aashok@gsu.edu}

\begin{abstract}
% \textcolor{red}{SK: Needs work. One of the main innovations is to realize that we can transmit information in shapes. Rice grain symbols is one way to do it. I will work on this at the very end.}
% \iffalse
% Revelio is a novel screen-camera communication system designed for real-world applications, enabling visually imperceptible data embedding. By leveraging temporal flicker fusion in the OKLAB color space and adaptive modulation techniques, Revelio ensures unobtrusive meta-information transmission. The encoder uses Reed-Solomon codes and rice grain symbols, while the decoder uses advanced neural networks for robust data extraction. Initial applications include preventing interactive television and video piracy. This work demonstrates the feasibility of reliable, real-time, and imperceptible communication via standard digital screens and smartphones.
% \fi
We present `Revelio', a real-world screen-camera communication system leveraging temporal flicker fusion in the OKLAB color space. Using spatially-adaptive flickering and encoding information in pixel region shapes, Revelio achieves visually imperceptible data embedding while remaining robust against noise, asynchronicity, and distortions in screen-camera channels, ensuring reliable decoding by standard smartphone cameras. The decoder, driven by a two-stage neural network, uses a weighted differential accumulator for precise frame detection and symbol recognition. Initial experiments demonstrate Revelio's effectiveness in interactive television, offering an unobtrusive method for meta-information transmission. %Initial experiments show significant improvements in imperceptibility and decoding accuracy, paving the way for interactive media and content security.
\end{abstract}


% \iffalse
% \begin{terms}
% \end{terms}
% \fi
\begin{figure*}[htb!]
\includegraphics[width=\textwidth]{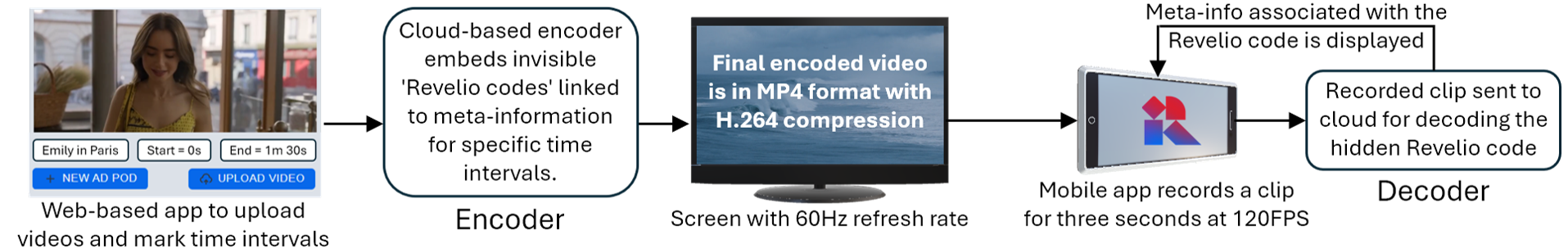}
\caption{Overall system: (i) A web app to upload videos and link meta-information to ``Revelio codes"; (ii) a cloud-based encoder that invisibly embeds these codes; (iii) a mobile app, with cloud decoding, that extracts the hidden data by recording the screen.}
% \Description[Overall System diagram of Revelio]{Overall System Diagram of Revelio - The Figure illustrates the overall system diagram of Revelio as a screen-camera communication system. Revelio has three main components: (i) A web-based app for creators to upload videos and link meta-information to Revelio codes; (ii) a near-real-time cloud-based encoder that embeds Revelio codes invisibly in videos; and (iii) a mobile app with cloud-based decoding, that lets viewers extract hidden data by pointing their phones at the screen.}
\vspace{-4mm}
\label{fig:system_diagram}
\end{figure*}

\section{Introduction}
Screen-camera communication adds interactivity to videos by embedding invisible data (meta-information), offering an unobtrusive yet enriched viewing experience~\cite{Morkel2005, Cox2007, Zhang2020, noisy,stegastamp, Wengrowski_2019_CVPR, Zhang2019, DeepLight2021,Woo2012,icassp, japanese_paper, chromacode, unseen-code, aircode}. Viewers can access hidden data by simply pointing their smartphones at the screen. Applications include interactive and shoppable television, as well as video piracy protection, where embedded data acts as tamper-evident markers making screen-grabbed content unusable, and helping in the identifying malicious users.

Data embedding techniques have advanced from classical steganography using domain transformations~\cite{Morkel2005, Cox2007, Zhang2020, noisy} to modern deep learning methods~\cite{stegastamp, Wengrowski_2019_CVPR, Zhang2019, DeepLight2021}. In this work, we use temporal flicker-fusion to embed data into videos that is invisible to the human eye. Flicker-fusion, either lightness-based~\cite{DeepLight2021, chromacode, aircode, inframe} or color-based~\cite{Woo2012, icassp, japanese_paper, unseen-code}, alternates pixel lightness or color between consecutive frames (e.g., lightness is changed by $+d$ in one frame and by $-d$ in the next). Rapid changes are averaged out by the human eye, preserving video quality, while a camera sampling at twice the encoding rate detects flickering. Fusion occurs when flickering is imperceptible to human eyes: around 60 Hz for lightness and 25 Hz for color~\cite{25HZ}\cite{60HZ}. If $d$ is too large, flickering is perceptible even at 60Hz. Hence controlling flickering strength is crucial, though reducing it can affect decoding. Additionally, opposite pixel changes may leave perceptible static residuals. We refer to perceptible flickering and static residuals as {\em visual artifacts}.

{\bf Design Goals and Main Contributions:}
We embed a fixed-length payload of bits, called a ``Revelio code", in a scene. Multiple scenes can have different Revelio codes, each independently linked to some meta-information such as authorship or second-screen experiences. A 48-bit code supports about 281 trillion distinct Revelio codes, enabling extensive data embedding. Rather than high data rates, we prioritize robust and unobtrusive communication of a modest payload for real-world applications.
Our goal is to have an encoder that can quickly embed data across various video types, scene dynamics, or frame rate with imperceptible visual artifacts on 60Hz screens. The decoder should extract hidden data quickly and reliably under typical well-lit conditions at 1.5 to 3 meters from the screen and varying angles, using a standard smartphone. Our key contributions include: (i) {\em Spatially-adaptive} flickering in OKLAB~\cite{OKLAB} on {\em all} the channels to minimize visual artifacts; (ii) Encoding data in the {\em shape} of pixel regions (symbols) rather than Manchester-style encoding; (iii) {\em Efficient packing} of symbols to improve decoding performance; (iv) Flickering the \emph{frame boundary} to identify reliably the region of interest; (v) A \emph{weighted cumulative differential} decoder, to reliably highlight the symbols and use {\em time diversity} for enhanced decoding; (vi) Two {\em neural networks}: one to identify the video frame and another to detect symbol shapes.

{\bf Related Work:}
While extensive research on flicker fusion techniques exists, we focus on closely related works~\cite{DeepLight2021, Woo2012, icassp, japanese_paper, chromacode, unseen-code, aircode, inframe}. However, these methods fall short of meeting our goals, as we briefly explain below. In~\cite{chromacode,  unseen-code, aircode, inframe}, lightness flickering requires low strength at 60Hz to minimize visual artifacts, with reliable decoding needing 120Hz rendering. In~\cite{DeepLight2021, icassp, Woo2012, japanese_paper, unseen-code}, color flickering is used, with~\cite{japanese_paper, unseen-code} using the XYZ space and~\cite{DeepLight2021} using the blue RGB channel. Since RGB and XYZ are not orthogonal~\cite{OKLAB} spaces, altering the color also affects the lightness, even at 60Hz a low flickering strength is needed to reduce visual artifacts. In~\cite{icassp, japanese_paper, chromacode, unseen-code, aircode, inframe}, Manchester-style encoding is used, 
%where bit 0 (1) is represented by changes of $+d$ and $-d$ ($-d$ and $+d$) across consecutive frames 
but it is less effective against the rolling shutter effect and screen-camera asynchronicity, which cause frame splitting or merging. Solutions such as inversion-invariant patterns~\cite{inframe}, least-squares method, and pilot symbols~\cite{icassp, chromacode} address these issues but reduce data rates or complicate decoding.
Methods that encode information in rectangular cells,  with binary patterns~\cite{DeepLight2021, Woo2012, japanese_paper, chromacode, unseen-code, aircode, inframe}, require precise screen detection for reliable decoding. In~\cite{inframe}, visible QR codes aid in detection, but are obtrusive to the viewer. In~\cite{chromacode}, clear screen borders are assumed and edge detection is used, which can pick up spurious edges. In~\cite{DeepLight2021, aircode}, the screen is detected rather than the frame, while~\cite{icassp}, uses a computationally intensive RANSAC algorithm to detect the region of interest. %In~\cite{aircode} visual odometry is used for robust screen tracking, increasing decoder complexity. 
Uniform flickering across pixels, common in prior work,
%in~\cite{japanese_paper} the color gamut is narrowed, 
%(where the flicker direction in the XYZ color space is fixed), 
can be improved with spatially-adaptive flickering, though~\cite{icassp, chromacode} flicker only {\em one} channel while keeping others constant. %In~\cite{chromacode}, lightness is flickered while chromaticity and hue remain constant, and in~\cite{icassp}, one color channel is flickered while the others are kept fixed. 

\section{System Architecture}

The overall system is shown in Fig.~\ref{fig:system_diagram}. For simplicity, we assume that a single Revelio code is embedded in the entire HD (1920x1080) video. 
%The encoder embeds a unique Revelio code into each interval linked to meta-information. 
Our mobile app extracts the code by sending a three-second, 120FPS
%~\footnote{Recording at 60FPS slightly reduces decoding performance, so we use 120FPS, which most smartphones support, for better results.} 
screen recording to the decoder. We assume that the screen has a 60Hz refresh rate. 
\vspace{-2mm}
\subsection{Encoder}
 Since no color space is perfectly orthogonal, modifying one or more color channels affects lightness, requiring flickering at 60Hz for proper fusion. Thus, videos are up-sampled to 60FPS using sample-and-hold (e.g., duplicating frames for videos at 30FPS). We encode as follows.
%To achieve our objectives, we make the following choices: (1) Flickering is done in a perceptually uniform space with orthogonal channels\footnote{Changing one channel minimally impacts the other two.} to balance imperceptibility of visual artifacts with decoding performance; (2) We avoid Manchester-style encoding, which adds complexity and is less effective against rolling shutter effect and asynchronicity of the screen-camera channel; (3) Frame detection is prioritized over screen detection since most videos don’t fully occupy the screen; (4) Spatially-adaptive flickering on {\em all channels} to minimize visual artifacts and ensure robust decoding. 

{\em Color space:} Experiments with RGB, CIELAB, XYZ, and OKLAB~\cite{OKLAB} showed that flickering in OKLAB produced the fewest visual artifacts while maintaining decoding performance. OKLAB, with channels $L$ (lightness), $A$ (red-green), and $B$ (blue-yellow), offers better perceptual uniformity and orthogonality than CIELAB~\cite{OKLAB}. We convert all original RGB frames to OKLAB using the functions in~\cite{OKLAB}.

{\em Embed Information in Shapes:} 
We create a 1920x1080 {\em data frame}, with all pixels set to 0, for each original frame. Pixels to be flickered in OKLAB are set to 1, with their strength determined later. We encode information in the shape of pixel regions (symbols) to handle asynchronicity and the rolling shutter effect. The data frame has a grid of 16x9 symbols, placed maximally apart and 60 pixels from the boundary. Each symbol is chosen from four shapes: ellipses angled at 0$^\circ$, 45$^\circ$, 90$^\circ$, and 135$^\circ$, with pixels set to 1. The 45$^\circ$ and 135$^\circ$ ellipses have major axes of 100 pixels, while the others are 74 pixels; all have a minor axis of 20 pixels. %These shapes ensure reliable decoding, even with reduced image quality at a larger distance. 
Each symbol maps to two bits, providing 288 bits for encoding a 16-bit Revelio code with a Reed-Solomon (RS) code using an 8-bit alphabet, shortened to 36 alphabets. A 13-pixel-wide border around the data frame is set to 1, which aids in {\em frame detection}. The data frames alternate, with all pixels set to +1 in one frame becoming -1 in the next.

\begin{figure}[H]
\includegraphics[width=\columnwidth]{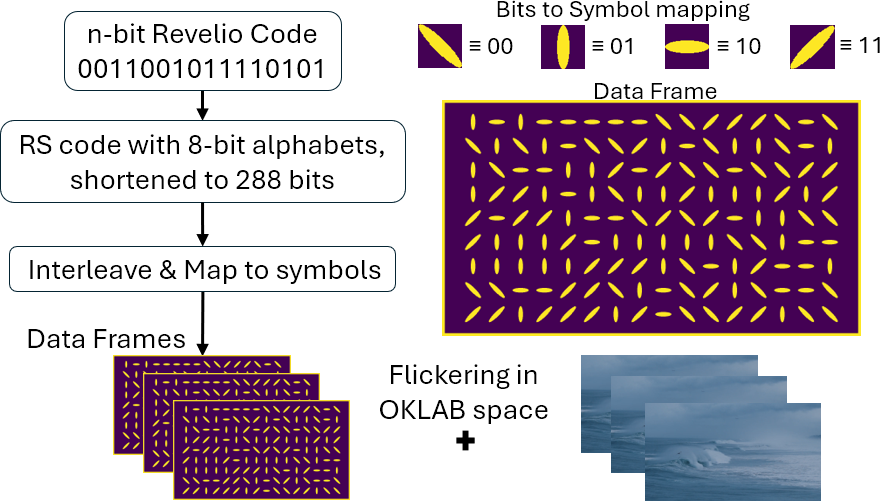}
\caption{Encoder pipeline: A 16-bit Revelio code is (Reed-Solomon) encoded to 288 bits, interleaved, mapped to a symbol, and flickered in the OKLAB space.}
% \Description[This diagram illustrates the Encoder]{Encoder  - This figure illustrates Revelio's encoder pipeline. A 16-bit Revelio code is Reed-Solomon encoded to 288 bits, interleaved, mapped to a rice grain, and flickered in the OKLAB space.}
\vspace{-4mm}
\label{fig:encoder}
\end{figure}

{\em Symbol Interleaving:} Four symbols (8 bits) of an RS alphabet are packed in a 2x2 grid, preserving the entire alphabet under favorable pixel conditions. %similar to water-filling in wireless communication~\cite{???}. 
Since there are nine rows, two RS alphabets are placed sequentially in the last row.

\begin{figure}[t!]
\centering
\includegraphics[width=0.75\columnwidth]{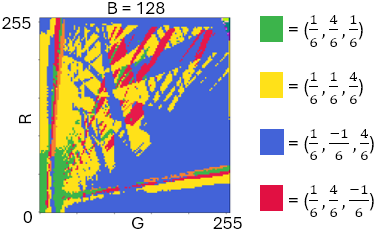}
\caption{Best $(\lambda, \alpha, \beta)$ values for $B = 128$ and all $R, G$.}
% \Description[This diagram illustrates a heatmap]{Heatmap  - This figure illustrates encoder heatmap for a fixed blue value of 128 and for all values of the red and the green colors.}
\label{fig:heatmap}
\vspace{-4mm}
\end{figure}

{\em Spatially-Adaptive Flickering of All Channels:} %When the data frame has a 1, the corresponding pixel in the original frame is adjusted in OKLAB. 
Finding the right flickering strength, $d$, is hard due to the difficulty in modeling visual artifacts. We proceed by fixing $d$ and splitting it across $L, A, B$ as $(\lambda d, \alpha d, \beta d)$, with $\vert\lambda\vert + \vert\alpha\vert + \vert\beta\vert = 1$. This roughly maintains a fixed decoding performance for different $\lambda, \alpha, \beta$.  %\footnote{For some bright or dark colors, it may be required to reduce the flickering strength to minimize visual artifacts.}. 
We use symmetric flickering to adjust the pixels by $(\lambda d, \alpha d, \beta d)$ in one frame and $(-\lambda d, -\alpha d, -\beta d)$ in the next. Although splitting minimizes visual artifacts, it reduces decoding performance, and increasing $d$ can offset this loss. We use a finite set of experimentally chosen $\lambda, \alpha, \beta$ values to minimize visual artifacts across all R,G,B values. Let $(r, g, b)$ be the original RGB pixel, and $(r_1, g_1, b_1)$ and $(r_2, g_2, b_2)$ the changed values in consecutive frames, after conversion to OKLAB and back to RGB. Since the eye is more sensitive to lightness~\cite{25HZ, 60HZ}, we minimize $\omega\big\vert r - \frac{r_1+r_2}{2} \big\vert + \chi\big\vert g - \frac{g_1+g_2}{2} \big\vert + \gamma\big\vert b - \frac{b_1+b_2}{2} \big\vert$, to keep the fused lightness close to the original.
%\footnote{The eye is more sensitive to lightness than color~\cite{25HZ, 60HZ}, so that green is weighted more.}. 
Based on experiments, we use $\omega = 0.27$, $\chi = 0.7$, $\gamma = 0.03$ for $0 \leq L \leq 0.95$, and $\omega = \chi = \gamma = \frac{1}{3}$ for $0.95 < L \leq 1$. See Fig.~\ref{fig:heatmap}. 

%We assume the human eye, at 60Hz, perceives the average color~\cite{???}, so we optimize, $\omega\big\vert r - \frac{r_1+r_2}{2} \big\vert + \chi\big\vert g - \frac{g_1+g_2}{2} \big\vert + \gamma\big\vert b - \frac{b_1+b_2}{2} \big\vert$, to keep the perceived lightness close to the original\footnote{The eye is more sensitive to lightness than color~\cite{25HZ, 60HZ}, so green is weighted more.}. Based on experiments, we use $\omega = 0.27$, $\chi = 0.7$, $\gamma = 0.03$ for $0 \leq L \leq 0.95$, and $\omega = \chi = \gamma = \frac{1}{3}$ for $0.95 < L \leq 1$. Figure~\ref{fig:heatmap} shows the best $(\lambda, \beta, \alpha)$ for $B = 128$ and all $R, G$. 

Parameters such as symbol count, size, and border width can be optimized for specific use cases. Higher data rates are possible by adding more shapes, though this reduces decoding performance. The objective function mirrors the luminance formula~\cite{itu_hdtv}, and exploring a CIELAB-like color difference formula~\cite{colorDiffFormula} for OKLAB could improve results. Expanding ($\lambda, \alpha, \beta$) may further reduce visual artifacts, while concatenated codes can enhance error correction. To improve decoding, we record at 120FPS (supported by most smartphones) instead of 60FPS, which slightly reduces performance.

\vspace{-2mm}
\subsection{Decoder}
The screen recording is split into frames, converted to OKLAB, and a random set of $N$ (typically 8 to 15) consecutive frames, $\{F_i\}_{i=1}^N$, is  processed, referred to as a decoding {\em epoch}.

\begin{figure}[t!]
\includegraphics[width=\columnwidth]{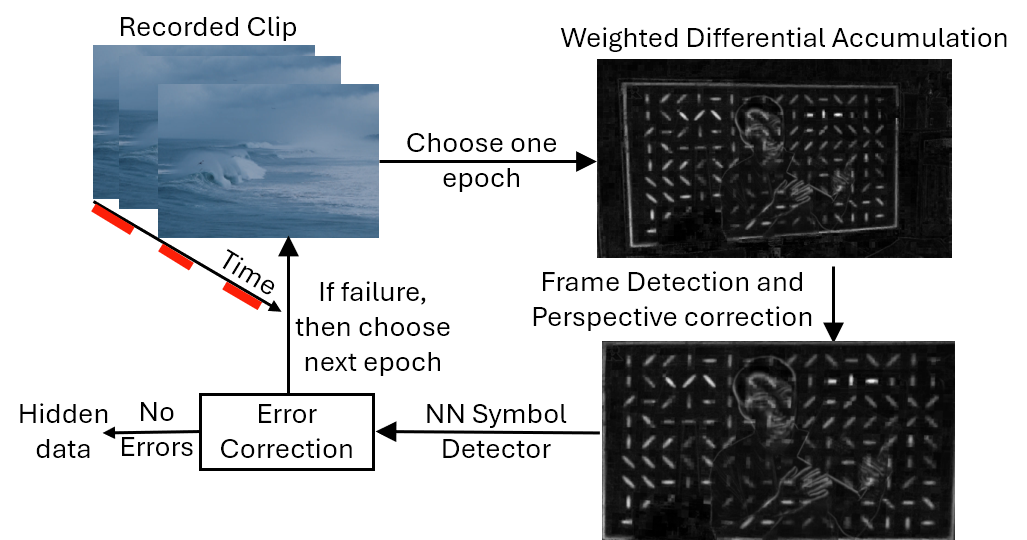}
\caption{Decoder pipeline: A decoding epoch processes $N$ (typically 8 to 15) consecutive frames.}
% \Description[This diagram illustrates the Decoder]{Decoder  - This figure illustrates Revelio's decoder.}
\vspace{-4mm}
\label{fig:decoder}
\end{figure}

% {\em Weighted Differential Accumulation in an Epoch:} In an epoch, we compute: $F^{L,A,B} = F^{L,A,B}_1 + \sum_{i = 1}^{N-1} w_i \big\vert F^{L,A,B}_i - F^{L,A,B}_{i+1}\big\vert$, for all channels, where $w_i$ decays with $i$, giving more importance to earlier frames, and then normalized to [0,255]. To minimize the effects of edges (in the original frame) on symbol detection, we apply Gaussian filtering to the $L$ channel before accumulation. To detect the region of interest and symbols, we simplify using the combined frame $F^A + F^B + cF^L$, where $c < 1$ reduces the influence of the $L$ channel. See Figure~\ref{fig:decoder}.
{\em Weighted Differential Accumulation in an Epoch:} In each epoch, we compute $F^{L,A,B} = F^{L,A,B}_1 + \sum_{i=1}^{N-1} w_i \big\vert F^{L,A,B}_i - F^{L,A,B}_{i+1}\big\vert$ for all channels, where $w_i$ decays with $i$, prioritizing earlier frames, and normalize the result to [0,255]. To reduce edge effects (from the original frame of the content) on symbol detection, we apply Gaussian filtering to the $L$ channel before accumulation. For region of interest and symbol detection, we simplify using the combined frame $F^A + F^B + cF^L$, where $c < 1$ limits the influence of the $L$ channel. See Fig.~\ref{fig:decoder}.

{\em Frame Detection and Perspective Correction:}
We use a Deeplab v3 Resnet18 neural network (NN)~\cite{deeplab_v3, resnet_kaiming_He} to segment the video frame in the above image, followed by Hough line transform for corner detection and perspective correction to 1920x1080. The NN is trained on thousands of images gathered from screens ranging from 27 to 65 inches, ensuring 80\% screen coverage in the camera preview without zooming. 
%Trained on thousands of images~\footnote{Training data was gathered with screen sizes from 27 to 65 inches, ensuring 80\% screen coverage in the camera preview without zooming.}, the NN segments the region of interest. The Hough line transform is then used to detect corners, followed by perspective correction to obtain a 1920x1080 frame. 

{\em Symbol Detection and Error-correction:}
The perspective-corrected frame is used to detect the shape of each of the 144 symbols. The decoder knows the location of each symbol's {\em patch}, a square, 15\% larger than the symbol's bounding box.  To account for frame detection errors, we take nine patches per symbol: one at the original location and eight jittered by a fixed amount, in the N, S, E, W, NE, NW, SE, SW directions. These patches are processed by a neural network to identify the shape or mark it as erased, with final decisions made by majority voting. Identified shapes are mapped to their corresponding bits, while erased symbols lead to bit erasures. The bit stream is de-interleaved and corrected by the RS decoder.

{\em Symbol Shape Detection:} 
We use a three-layer ResNet~\cite{resnet_kaiming_He} to quickly detect symbol shapes across 1,296 patches ($144 \times 9$). The NN learns to classify each patch into one of four shapes using perspective-corrected frames. During training, only the original patch for each symbol is cropped and labeled with its known
%\footnote{For training, the ground truth data is known at the decoder} 
shape, as the decoder has ground truth data. %(e.g., $0$, $1$, $2$, $3$ for $45^\circ$, $0^\circ$, $90^\circ$, $135^\circ$). 
During inference, if the probability difference between the top two softmax predictions exceeds a threshold (e.g., 0.35), the top class is chosen; otherwise, the patch is marked as an erasure. Although the 4-class NN can misclassify symbols, training with many diverse patches helps mark unclear symbols as erasures, which RS codes handle better than errors.
%See figure~\ref{fig:logits}.
%\begin{figure}[H]
%\includegraphics[width=\columnwidth]{Figures/Logits Picture.png}
%\caption{Logits of 4 class NN symbol Decoder for different scenarios. Notice that the logits are flat for the erasure case.}
% \Description[This diagram illustrates Logits of 4 class Neural Network symbol Decoder]{This figure depicts the logits of the Neural Decoder for different scenarios. Notice that the logits are flat for the erasure case.}
%\label{fig:logits}
%\end{figure}

{\em Time Diversity:} To account for hand movement and varying conditions, we use three disjoint decoding epochs sequentially. If two or more epochs fail, we combine their bit streams, resolve erasures with non-erased bits, mark conflicts as erasures, and then apply RS decoding for better performance.

\section{Evaluation}

\begin{figure}[t!]
\centering
\includegraphics[width=0.75\columnwidth]{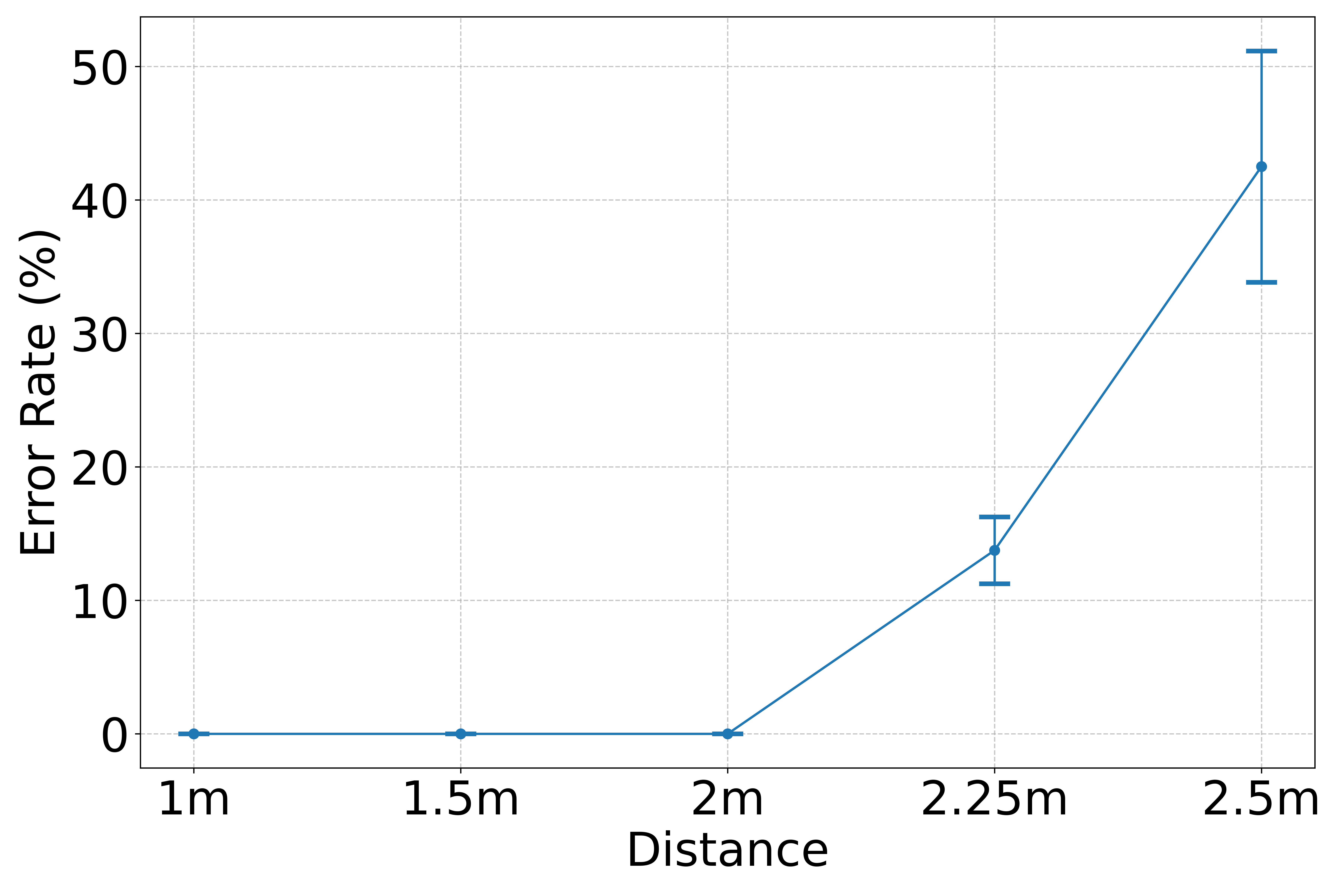}
\caption{Error Rate vs. Distance}
\label{fig:dis_error}
\vspace{-4mm}
\end{figure}

\begin{figure}[t!]
\centering
\includegraphics[width=0.95\columnwidth]{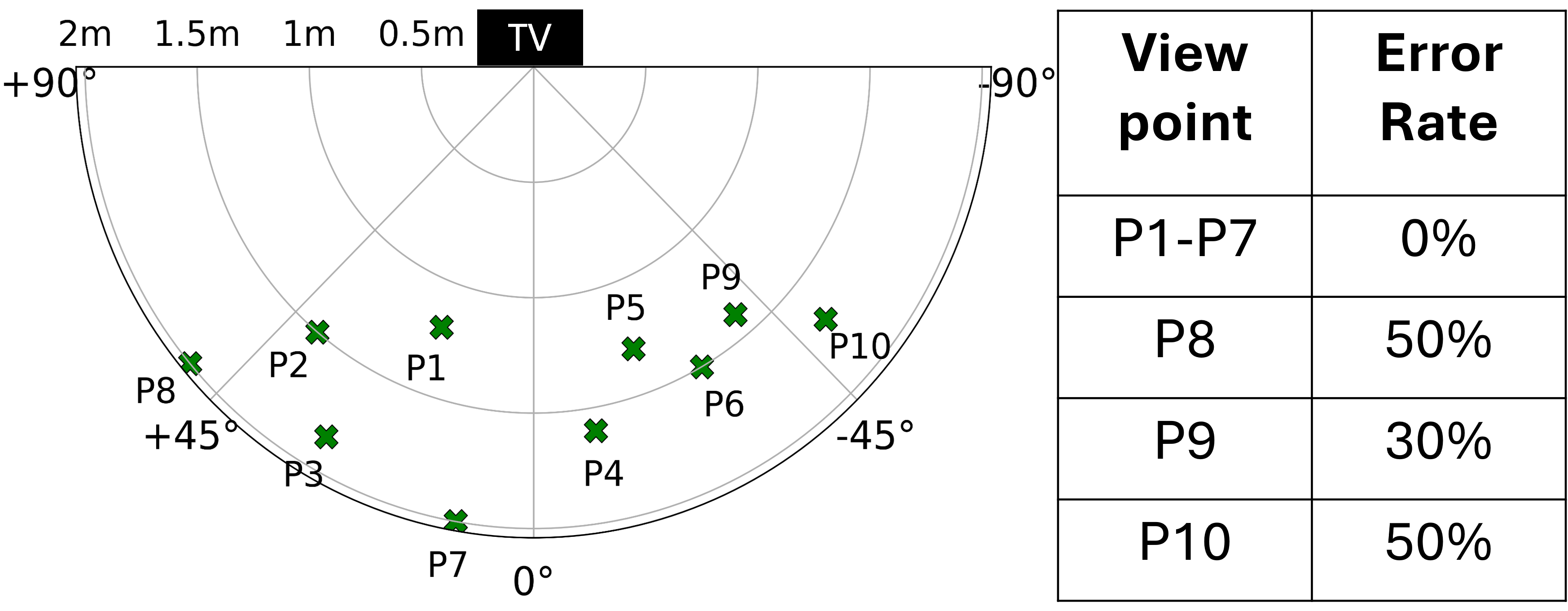}
\caption{Error Rate vs. View Points}
\label{fig:rand_error}
\vspace{-4mm}
\end{figure}

\begin{table}[h]
\centering

\scalebox{0.75}{
\begin{tabular}{|c|c|c|c|c|c|}
\hline
Video & Metric & \multicolumn{2}{c|}{d = 0.0375} & \multicolumn{2}{c|}{d = 0.0425} \\
\cline{3-6}
 & & Avg & Std & Avg & Std \\
\hline
\multirow{3}{*}{Big Buck Bunny} & VMAF & 97.045 & 1.465 & 96.879 & 1.575 \\
 & PSNR & 40.826 & 1.106 & 40.039 & 1.003 \\
 & SSIM & 0.981 & 0.007 & 0.979 & 0.008 \\
\hline
\multirow{3}{*}{Emily in Paris} & VMAF & 95.586 & 0.439 & 95.39 & 0.432 \\
 & PSNR & 42.776 & 0.579 & 41.84 & 0.585 \\
 & SSIM & 0.985 & 0.002 & 0.982 & 0.002 \\
\hline
\multirow{3}{*}{Uncle Roger} & VMAF & 95.629 & 0.612 & 95.574 & 0.597 \\
 & PSNR & 41.726 & 1.049 & 40.706 & 1.035 \\
 & SSIM & 0.986 & 0.003 & 0.984 & 0.003 \\
\hline
\multirow{3}{*}{Skyfall} & VMAF & 95.997 & 0.391 & 95.863 & 0.437 \\
 & PSNR & 44.626 & 1.333 & 43.92 & 1.537 \\
 & SSIM & 0.989 & 0.003 & 0.988 & 0.003 \\
\hline
\end{tabular}
}
\caption{Video quality scores. VMAF $>$ 95, SSIM $\approx 1$, PSNR $>$ 40 dB, are considered excellent.}
\vspace{-4mm}
\label{tab:video_metrics_encoding_depth}
\end{table}

\noindent{\bf Metric:} A system run is deemed success when the 16-bit Revelio code payload is correctly decoded at the mobile (camera) device. We use the \emph{Error Rate} metric, which reflects the percentage of the failures in the system across all system runs. We also evaluate the quality of the encoded videos using standard metrics: VMAF (Video Multi-Method Assessment Fusion)~\cite{vmaf}, PSNR~\cite{psnr_ssim}, and SSIM~\cite{psnr_ssim}. 

\vspace{1mm}\noindent{\bf Setup and Method:} 
We evalaute the performance of our system through the \emph{Error Rate} metric. We conducted experiments on a 65-inch 4K UHD LG TV set to the `Vivid' mode, with `Motion Smoothing,' `Super-Resolution,' and color corrections disabled to minimize visual artifacts. Each 1920x1080 video was encoded using a different Revelio code, with a flickering strength of $d = 0.0425$, ensuring {\em imperceptible} data embedding. The recordings were made using our app on an iPhone 15 Pro Max in both day and night ambient lighting. We evaluated decoder performance across various view points (distance and angle from the screen) and video types, including `Big Buck' (animated, 60FPS), `Emily in Paris -- Netflix' (natural, 30FPS), `Uncle Roger -- YouTube'
%~\footnote{`Emily in Paris' is on Netflix; `Uncle Roger' is on YouTube.} 
(natural, 24FPS), and a dark scene from `Skyfall' (movie, 24FPS). Zoom was disabled. The system was tested in a \emph{hand-held} setup by two users while \emph{seated}, maintaining a steady position.

%By integrating these results from these experiments, we aim to understand the limitations and strengths of the system in practical deployments. 

\textbf{Error Rate vs. Distance:}
We evaluate the system with the phone pointed directly at the screen (0$^\circ$ angle) at distances of 1m, 1.5m, 2m, 2.25m, and 2.5m. The marker represents the mean value with the minimum and maximum depicted by the whiskers. 
Fig.~\ref{fig:dis_error} shows that up to 2m, the system consistently maintained zero errors on all videos, indicating extremely reliable data transmission. At 2.25m and 2.5m, error rates increased, and we identified a lot of erasures as the screen frame was not detected accurately resulting in perspective artifacts on the encoded pixels resulting in erasures while decoding. 

\textbf{Error Rate vs. View Points:} We evaluated the system at different angles, using ten random view points (P1 to P10 in Figure~\ref{fig:rand_error}) at varying distances and angles ($0^\circ$ being the direct line of sight). Multiple experiments were conducted at each point and videos were randomly selected for each trial. From the Fig.~\ref{fig:rand_error}, we can observe that the points P1 to P7, located at distances between 1m and 2m with moderate angles of $20^\circ$ to $40^\circ$ showed the lowest error rates. P8, P9 and P10, with angles greater than $40^\circ$, showed higher error rates, all attributed to frame detection failures.

\textbf{Remarks:} Our experiments show that the system performs well up to 2m and angles of $\pm40^\circ$ across various videos. Beyond 2.25m and larger angles, error rates can be reduced by (i) enabling camera zoom in our app, (ii) training the neural network for frame detection at greater distances and angles, and (iii) refining Hough line transform thresholds. The decoder typically succeeded in the first epoch, finishing processing within 6 seconds after uploading the recorded clip. In a handful of cases, additional epochs were needed, highlighting the value of time diversity, especially at greater distances and angles.
Table~\ref{tab:video_metrics_encoding_depth} shows that the original (excellent scores) video quality is well preserved by our system. 

%From these experiments, we conclude that the errors that occur in 2.5m are attributed towards screen detection errors which can be circumvented by zooming. We can considerably use a 4X digital zoom which not only increases the reliable distance of our system up to 8 meters, and for a 1080p stream, we can have reliable quality in the video without losing any information on the code. Also, we find that this is where time diversity and repetitive coding come in, which helps in boosting the reliability of the system. The random point experiment also shows us that our system works well up to angles of $\pm40^\circ$, which can be further enhanced by training on more complex scenarios. We also find the current hough line transform thresholds can be improved further to work at farther distances, and we plan to improve this by using a neural system that will do the corner detection of the screen for perspective warping.
\section{Conclusion}
We presented a novel system for embedding data in videos by encoding information in symbol shapes with spatially-adaptive flickering across all channels in OKLAB. This balances imperceptibility and performance. A lightweight ResNet detects symbol shapes, while another neural network identifies the region of interest. Reed-Solomon error correction and multiple decoding epochs ensure reliable decoding under challenging conditions. Designed for efficiency and accuracy, the initial results show the potential of Revelio for interactive television.

A key challenge is maintaining imperceptibility for very dark/bright tones. Future work can focus on an optimization problem to balance visual artifacts with decoding performance across the entire RGB space, using suitable cost functions and relaxation techniques. Developing a robust metric for imperceptibility is crucial. In our system, 48-bit payload is possible with reduced decoding performance. Increasing the bit rate without losing performance is a key research area. A 4x digital zoom can extend the range to 8 meters, and superimposition coding~\cite{sup_coding} can transmit larger payloads to nearby viewers and smaller ones to those farther away.

\newpage
% \footnotesize
\balance
\bibliographystyle{IEEEbib}
\bibliography{references}

%\newpage
% \appendix

\end{document}